\documentclass[a4paper,11pt,preprintnumbers]{article}
\pdfoutput=1 

\usepackage{jcappub} 

\usepackage[T1]{fontenc} 

\usepackage[normalem]{ulem}

\title{Suppression of Fast Neutrino Flavor Conversions Occurring
at Large Distances in Core-Collapse Supernovae}

\author[a]{Sajad Abbar,}
\author[b,c]{Francesco Capozzi}

\affiliation[a]{Max-Planck-Institut f\"ur Physik (Werner-Heisenberg-Institut),\\ F\"ohringer Ring 6, 80805 M\"unchen, Germany}
\affiliation[b]{Center for Neutrino Physics, Department of Physics, Virginia Tech, Blacksburg, VA 24061, USA}
\affiliation[c]{Instituto de F\'isica Corpuscular, Universidad de Valencia \& CSIC, Edificio Institutos de Investigaci\'on, Calle Catedr\'atico Jos\'e Beltr\'an 2, 46980 Paterna, Spain}

\emailAdd{abbar@mpp.mpg.de}
\emailAdd{fcapozzi@ific.uv.es}

\abstract{
Neutrinos propagating in dense neutrino media
such as  core-collapse supernovae
and neutron star merger remnants
can experience the so-called \textit{fast} flavor
conversions on scales
much shorter than those expected in vacuum. 
A very generic class of fast flavor  instabilities is  the ones which are produced by the backward scattering
of neutrinos off the nuclei at relatively large distances from the supernova core.
In this study we demonstrate  that despite their ubiquity, such fast instabilities are unlikely 
to cause significant flavor conversions 
if the population of neutrinos in the backward direction is not large enough.
Indeed, the scattering-induced  instabilities  can  mostly  impact  the 
neutrinos traveling in the backward direction, which represent only a small fraction of neutrinos at large radii.  
We show that this can be explained by the shape
of the unstable flavor eigenstates, which can be extremely peaked at the backward angles. 
}

\begin{document}
\begin{flushright}
MPP-2021-195\\
IFIC/21-50
\end{flushright}
\maketitle
\flushbottom


\section{Introduction}

Neutrino emission is a major effect in the most extreme astrophysical settings, such as 
core-collapse supernova (CCSN) explosions. The density of neutrinos in such environments
 is so high  
that the neutrino  self-interactions can play an important role in their flavor evolution. 
In particular, it leads to a nonlinear and rich phenomenon in which the whole
neutrino gas can oscillate collectively~\cite{Pastor:2002we,duan:2006an, duan:2006jv,duan:2010bg,  Chakraborty:2016yeg, Qian:1996xt, Duan:2015cqa}.

Collective neutrino oscillations can influence the physics of CCSNe 
by modifying  the spectra of  neutrinos and antineutrinos.
 On the one hand, they can influence the nucleosynthesis of heavy elements 
 by  shifting  the neutron-to-proton ratio. On the other hand,
 they can affect the SN dynamics by changing the neutrino energy deposition into
the stalling shock wave. 
Apart from these, they can also modify the  SN neutrino signal detectable
from a galactic CCSN.

The first studies on collective neutrino oscillations were centered around what we now call  \textit{slow} flavor conversions.
Slow modes occur on scales  determined by the neutrino vacuum frequency, $\omega = \Delta m^2/2E$,  which are
$\sim \mathcal{O}(1)$~km for a typical SN neutrino energy and atmospheric mass
splitting. 
 However, it was then perceived that 
a dense neutrino  gas can also experience \textit{fast} flavor conversions on extremely short 
scales~\cite{Sawyer:2005jk, Sawyer:2015dsa,
 Chakraborty:2016lct, Izaguirre:2016gsx, Wu:2017qpc, 
  Capozzi:2017gqd, Richers:2019grc,   
  Dasgupta:2016dbv, Abbar:2017pkh, Abbar:2018beu, Capozzi:2018clo, Martin:2019gxb, Capozzi:2019lso, Doring:2019axc, Chakraborty:2019wxe, Johns:2019izj, Cherry:2019vkv, Dasgupta:2021gfs, Bhattacharyya:2020jpj, Martin:2021xyl, Duan:2021woc, Li:2021vqj, Tamborra:2020cul, Wu:2021uvt, Sigl:2021tmj, Kato:2021cjf,  Richers:2021nbx, Morinaga:2021vmc, Nagakura:2021hyb, Richers:2021xtf, Sasaki:2021zld, Padilla-Gay:2021haz, Abbar:2020qpi, Capozzi:2020syn, DelfanAzari:2020fmq, Morinaga:2019apg, DelfanAzari:2019epo, Harada:2021ata, Kato:2020hlc}.  
Unlike the slow modes, the fast conversions scale as  
 $\sim G_{\rm{F}}^{-1} n_{\nu}^{-1}$,
with $n_\nu$ and $G_{\rm{F}}$ being the neutrino number density and the 
Fermi coupling constant, respectively.

The necessary and  sufficient condition 
for the occurrence of fast flavor instabilities is that 
the angular distribution of the neutrino electron lepton number (ELN) defined as,
\begin{equation}
  G_\mathbf{v} =
  \sqrt2 G_{\mathrm{F}}
  \int_0^\infty \frac{E_\nu^2 \mathrm{d} E_\nu}{(2\pi)^3}
        [f_{\nu_e}(E_\nu,\mathbf{v})- f_{\bar\nu_e}(E_\nu,\mathbf{v})],
\end{equation}
changes its sign at some directions~\cite{Morinaga:2021vmc}. In other words, fast instabilities exist provided that the  ELN angular distribution
 possesses a crossing.
Here $E_\nu$, $\mathbf{p}$, and $\mathbf{v}$
are the neutrino energy, momentum, and velocity, respectively,
and $f_\nu$'s are the neutrino occupation numbers with $G_{\mathrm{F}}$ being the Fermi constant.
Note that here we assume   $f_{\nu_{\mu}}=f_{\nu_\tau}=f_{\bar{\nu}_{\mu}}=f_{\bar{\nu}_\tau}$, otherwise
there can exist  crossings in the other leptonic channels, which can similarly lead to fast conversion 
modes~\cite{Capozzi:2020syn, Capozzi:2018clo}.

It has been demonstrated that ELN crossings and their associated fast instabilities can occur  in  different  SN regions
via different mechanisms. The deepest region where fast instabilities can appear is within 
the convective layer of the porto-neutron star (PNS),  well below the neutrinospheres of neutrinos and antineutrinos. 
These  instabilities occur due to
a shallow crossing between the angular distributions of $\nu_e$ and $\bar\nu_e$, which are 
 almost isotropic~\cite{DelfanAzari:2019tez, Abbar:2019zoq,Glas:2019ijo}.
 As a matter of fact,  large amplitude modulations in the spatial distributions of $\nu_e$ and $\bar\nu_e$
number densities can be caused by the strong convective activities. This implies that deep SN zones can always exist 
for which the number densities of $\nu_e$ and $\bar\nu_e$ can be extremely close to each other  and
an ELN crossing can occur regardless of the isotropic nature of neutrino angular distributions.
Though fascinating, 
the nearly equal distributions of neutrinos and antineutrinos of all flavors 
prevents a significant amount of flavor conversion resulted from such fast instabilities.

ELN crossings and their associated  fast instabilities can also occur 
within the neutrino decoupling region,  just above the PNS~\cite{Abbar:2018shq,Nagakura:2019sig,Abbar:2019zoq, Nagakura:2021hyb}.  
In the SN zones above the neutrinospheres, the  angular distributions of $\bar\nu_e$
 are more peaked in the forward direction than those of $\nu_e$, due to 
 their decoupling  at  smaller radii. This leads to a high chance of the occurrence of  ELN crossings 
in the  zones where $\nu_e-\bar\nu_e$ asymmetry is small~\cite{Abbar:2018shq, Shalgar:2019kzy}. 
Given this observation,  the asymmetry in the neutrino emission
  caused by   LESA (lepton-emission self-sustained asymmetry)~\cite{Tamborra:2014aua, Glas:2018vcs}
  plays a pivotal role in this context (see also Ref.~\cite{Nagakura:2019evv} for other relevant effects).

Finally, a very generic class of ELN crossings and fast instabilities can  occur farther away from the SN core. 
Though they were first discovered 
in the pre-shock SN region~\cite{Morinaga:2019wsv}, it was then understood that they
 can  occur in the post-shock region, as well~\cite{Abbar:2020qpi, Capozzi:2020syn}.
This type of ELN crossings is generated
 by the  backward scatterings of neutrinos off heavy nuclei or nucleons. 
 Generally speaking, $\bar{\nu}_e$ has a larger cross section than  ${\nu}_e$.
 This implies that at larger distances where the backward direction is almost empty,
 $\bar{\nu}_e$ can become the dominant type due to the  higher scattering rate and an ELN crossing can occur.
 
 In the present work, we study neutrino flavor conversions due to such scattering-induced fast instabilities.
 Despite being ubiquitous at large distances from the SN core, we show that they can not, in general, lead
 to significant neutrino  flavor conversions if the population of neutrinos in the backward direction
 is not large enough. 
 Our results are consistent with the ones provided in Ref.~\cite{Zaizen:2021wwl},
 which appeared while this work was under preparation. However, we here take a step forward
 and 
 give an explanation of this observation based on  the shape of  unstable eigenstates 
in the linear regime. We also provide a rough estimate of the depth of the ELN crossing below which
significant flavor conversions are suppressed. 
 
 The paper is organized as follows. In Sec.~\ref{sec:model}, we   introduce our model and 
  present our results of the linear stability analysis and we develop some useful insights regarding
  the form of the unstable wave functions. 
   We then discuss  neutrino flavor evolution in the nonlinear regime before the conclusion.

\section{Time-dependent one-dimensional neutrino gas model}\label{sec:model}
In this section, we discuss our time-dependent one-dimensional (1D) neutrino gas and we present our results
in the linear and nonlinear regimes, respectively. 

We consider a time-dependent gas of neutrinos, whose conditions are axially symmetric around the $z$ axis. We set to zero the neutrino mass splittings so that the evolution of the system is energy-independent. We also neglect the matter potential due to neutrino coherent forward scattering on electrons and also neutrino inelastic collisions. Under these conditions, the flavor evolution of the neutrino gas is governed by the following Liouville-von Neumann equation~\cite{Sigl:1992fn,Strack:2005ux,Cardall:2007zw,Volpe:2013jgr, Vlasenko:2013fja} ($c=\hbar=1$)

\begin{equation}
i (\partial_t + v \partial_z)
\varrho_{v} = \left[
   \mathsf{H}_{\nu \nu}^{v} ,
  \varrho_{v}\right]\,,
\label{Eq:EOM}
\end{equation}
where $\varrho_{v}(t, z)$ represents the usual density matrix for neutrinos with velocity $v$ along the $z$ axis, and
\begin{equation}
  \mathsf{H}_{\nu \nu}^v = \sqrt2 G_{\mathrm{F}}
  \int_{-1}^{+1}\!  \mathrm{d}v^\prime
  ( 1- v\,v^\prime)
  (\varrho_{v^\prime} - \bar\varrho_{v^\prime}).
\end{equation}
is the potential stemming from the neutrino-neutrino forward scattering
\cite{Fuller:1987aa,Notzold:1988kx,Pantaleone:1992xh}.

\subsection{Flavor instabilities in the linear regime}

In the two-flavor scenario ($e$ and $x$), the density matrix can be written as~\cite{Banerjee:2011fj},
\begin{align}
  \varrho = \frac{f_{\nu_e} + f_{\nu_x}}{2}
  + \frac{f_{\nu_e} - f_{\nu_x}}{2} \begin{bmatrix}
    s & S \\ S^* & -s \end{bmatrix},
\end{align}
in the weak-interaction basis
where  the complex and real scalar fields $S$ and
$s$ describe the flavor coherence and the flavor conversion of the
neutrino gas, respectively. 
 
  In the linear regime
  before any  significant flavor conversions occur, one has $|S_v|\ll1$ and
$s\approx 1$,  and Eq.~\eqref{Eq:EOM} can be written as~\cite{Banerjee:2011fj, Vaananen:2013qja, Izaguirre:2016gsx},
\begin{equation}
 i (\partial_t + v \partial_z) S_{v}
 = (\epsilon_0 +v\epsilon) S_v
 - \int\!\mathrm{d}v^\prime (1 - v\,v^\prime)
 G_{v^\prime} S_{v^\prime},
 \label{Eq:linear}
 \end{equation}
 where  $\epsilon_0 = \int\!\mathrm{d}v^\prime G_{v^\prime}$,
  $\epsilon = \int\!\mathrm{d}v^\prime G_{v^\prime} v^\prime$, and we have only retained the terms linear in $S_v$.
We look for plane wave solutions obeying
\begin{align}
  S_v(t,z) = Q_v\,  e^{-i\Omega t +
    iKz}\,.
\end{align}
An eigenmode of flavor conversion can be identified by its frequency and wavevector, $\Omega$ and $K$, respectively.

The neutrino gas is stable in the flavor space and 
the flavor mixing
amplitude $S_v$ remains small unless 
for a real wave vector
$K$, the corresponding $\Omega$ has a positive imaginary
component, i.e.\
$\Omega_\mathrm{i}=\mathrm{Im}(\Omega) >0$, referred  to as temporal instabilities which
is also the focus of this study. 
If such a condition is satisfied, $S_v$ can grow exponentially
and the neutrino medium can, in principle, experience non-negligible  flavor conversions. 
As discussed previously, such  instabilities 
exist if, and only if, the angular distribution of ELN possesses a crossing.

\begin{figure*}[tb!] 
\centering
\begin{center}
\includegraphics*[width=.49\textwidth, trim= 5 10 10 10,clip]{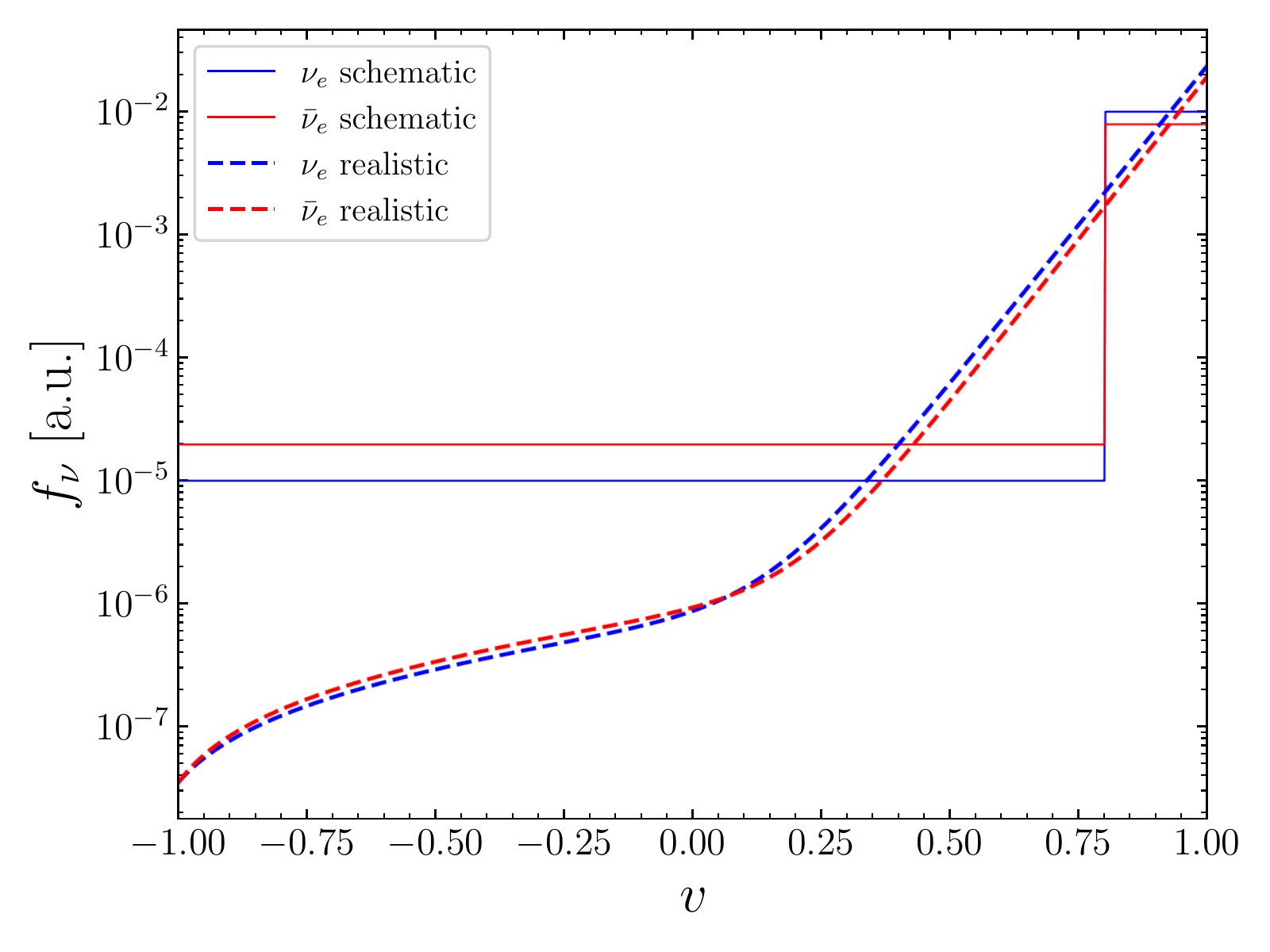}
\end{center}
\caption{
Neutrino angular distributions of realistic (dashed lines) and schematic (solid lines) cases where we have shown the  $\delta = 10^{-3}$ and $v_c=0.8$ case as an example.
}
\label{fig:f}
\end{figure*}

\begin{figure*}[tb!] 
\centering
\begin{center}
\includegraphics*[width=.99\textwidth, trim= 5 10 10 10,clip]{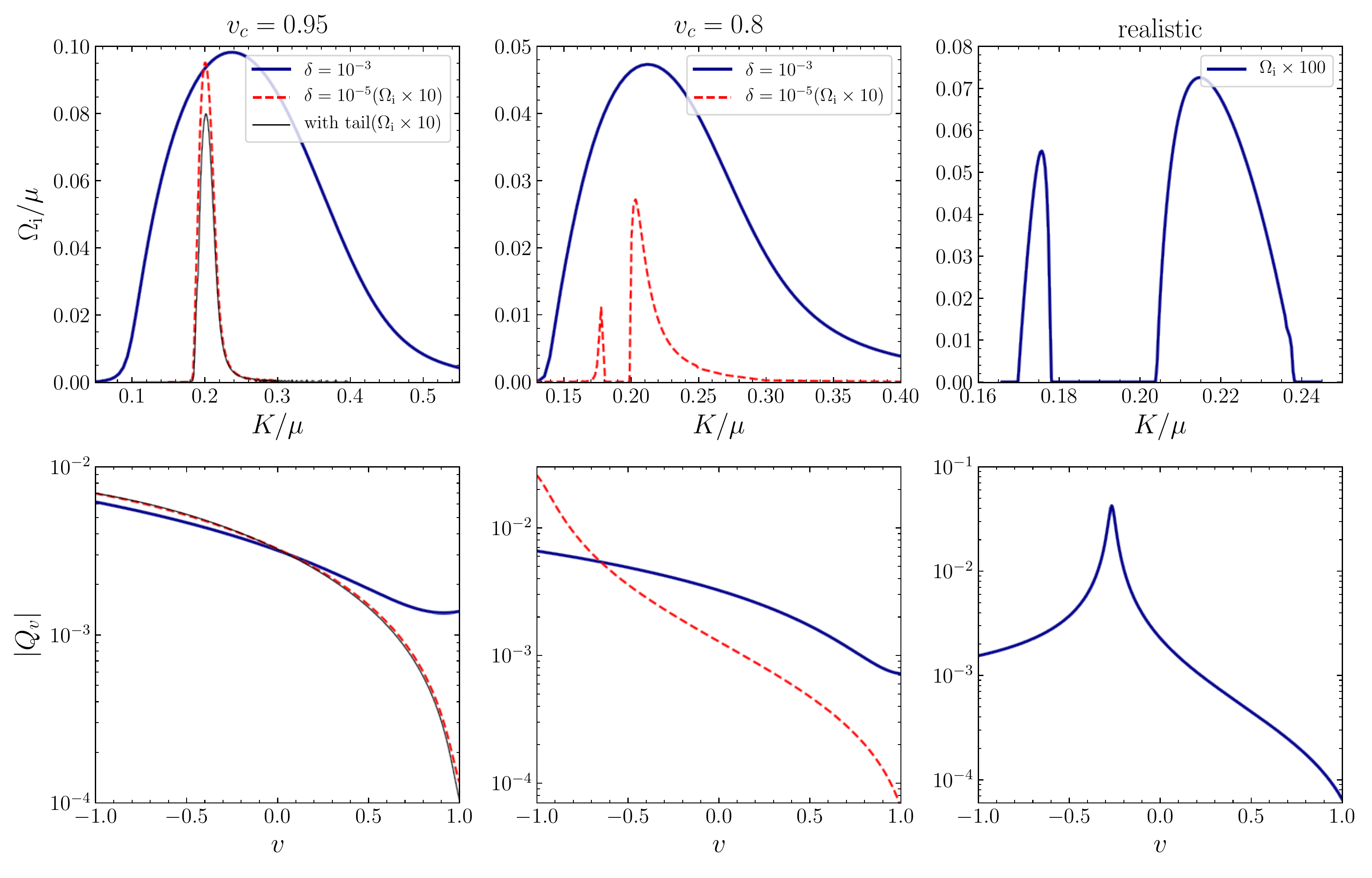}
\end{center}
\caption{
Upper panels: Growth rate of temporal fast instabilities 
       as functions of the real wave number $K$. 
       Lower panel: The  normalized angular distributions of the  eigenvectors, $Q_v$,
of unstable modes with maximum growth rate.
Note that the amplitude of $Q_v$ can be extremely small in the forward
direction where the majority of neutrinos are traveling.
This implies that the backward instabilities  should mainly affect the backward traveling neutrinos,
which are a small fraction of neutrinos at large radii.
The left and middle panels show the results obtained form our schematic ELN distribution (Eq.~(\ref{eq:schem})), while the
right panels are the ones from the realistic distribution (Eq.~(\ref{Eq:fnu})). 
For the sake of completeness, in the left panels we also study an ELN distribution with two crossings
       where the second  crossing exists in the tail of the  distribution, namely for $v<-0.9$ we have 
       a second crossing for which $G(v) = +\delta$.
}
\label{fig:linear}
\end{figure*}

In the following, we focus on two special classes of angular distributions which mimic crossings in the backward direction. 
We first consider a pure $\nu_e-\bar\nu_e$ neutrino gas with schematic distributions of the form:
  \begin{equation}
\label{eq:schem}
f_{\nu_e} \propto \left\lbrace
                \begin{array}{ll}
                  1\quad  \text{for}\quad v >v_c,\\
                   \delta \quad \text{for}\quad v<v_c.
                \end{array}
              \right.
\,\,    
, 
\quad
\quad     
f_{\bar\nu_e} \propto \left\lbrace
                \begin{array}{ll}
                  1 \quad \ \text{for}\quad v >v_c,\\
                   2\delta \quad \text{for}\quad v<v_c.
                \end{array}
              \right.
\,\,
\end{equation}
Here, $v_c$ is the point where the  crossing occurs, and $\delta$ is the depth of the backward 
scattering.
We also consider 
the more realistic neutrino angular distributions considered in Ref.~\cite{Zaizen:2021wwl},
\begin{equation}
f_\nu \propto g_b + g^\prime_{b} (e^{v+1}-1) + g_f b^{v-1},
\label{Eq:fnu}
\end{equation}
where the parameters $g_b$, $g^\prime_{b}$, $g_f$,  and $b$ can be different for $\nu_e$ and $\bar\nu_e$
and are presented in Table~I of Ref.~\cite{Zaizen:2021wwl}. Note that we here only take the normalized distributions from Eqs.~(\ref{eq:schem}) and (\ref{Eq:fnu})
with fixed $n_{\bar\nu_e}/n_{\nu_e} = 0.8$, though the results are independent of this choice. 
The neutrino distributions are indicated in Fig.~\ref{fig:f}

In the upper panels of Fig.~\ref{fig:linear}, we  indicate the  growth
rate of temporal fast instabilities, $\Omega_\mathrm{i}$, for a number of ELN distributions introduced in Eqs.~(\ref{eq:schem}) and (\ref{Eq:fnu}).
Here, we use
\begin{equation}
\mu = \sqrt2 G_{\rm{F}} n_{\nu_e},
\end{equation}
as a measure of the strength of the weak interactions.
Although  ELN crossings in the backward direction can be  extremely shallow, their corresponding
fast growth rates may not be necessarily small~\cite{Abbar:2020qpi}. 
For instance, for $\mu = 10^4$ km$^{-1}$ and $\delta=10^{-3}$, the growth rate can be as high as $~10^3$ km$^{-1}$,
which is  much larger than the typical growth rates expected for slow modes. 

In spite of such significant growth rates, 
fast instabilities brought about by backward ELN crossings
can not induce significant neutrino flavor conversions if the population of backward traveling
neutrinos is not large enough.
This can be deduced from the lower panels of Fig.~\ref{fig:linear},
where the  eigenvectors of  the unstable  modes
with maximum growth rates are shown. As can be clearly seen,
$Q_v$
 is highly peaked in the backward direction. 
This implies   that  the instabilities 
can mainly impact  the backward traveling neutrinos, which can be a tiny fraction of all neutrinos (at large radii). 


\subsection{Flavor conversions in the nonlinear regime}

We solve numerically the partial differential equation system in Eq.~(\ref{Eq:EOM}), assuming periodic boundary conditions. We calculate the spatial derivatives using the fast fourier transform based differentiation, whereas we solve temporal evolution using the backward differentiation formulae through the NAG library. Our initial conditions for the diagonal elements of $\varrho$ are independent of $z$,
 \begin{equation}
 \label{eq:schem2}
\varrho_{v,ee}(0,z) = \xi \times \left\lbrace
                \begin{array}{ll}
                  1\quad \text{for}\quad v >v_c,\\
                   \delta \quad \text{for}\quad v<v_c.
                \end{array}
              \right.
\,\,            
\bar{\varrho}_{v,ee}(0,z) = \xi \times \left\lbrace
                \begin{array}{ll}
                  0.8\quad \text{for}\quad v >v_c,\\
                   2\delta \quad \text{for}\quad v<v_c.
                \end{array}
              \right.
\,\,
\varrho_{v,xx}(0,z)=0,
\end{equation}
which, considering $\alpha=0.8$, allow one to reproduce $G_v$  in Eq.~(\ref{eq:schem}) with $\xi=4\times10^5$ km$^{-1}$. 
Note that $\mu \simeq (1-v_c)\xi$ for small $\delta$'s.
In terms of initial conditions for the off-diagonal elements, we choose them to be $v$-independent and we set $\varrho_{ex}(z)=\delta(z)$. This choice allows to give the same seed to all the $S_{v,K}(t)$ defined below and provides the necessary perturbation triggering fast flavor conversions.  

\begin{figure*}[tb!] 
\centering
\begin{center}
\includegraphics[width=0.49\textwidth]{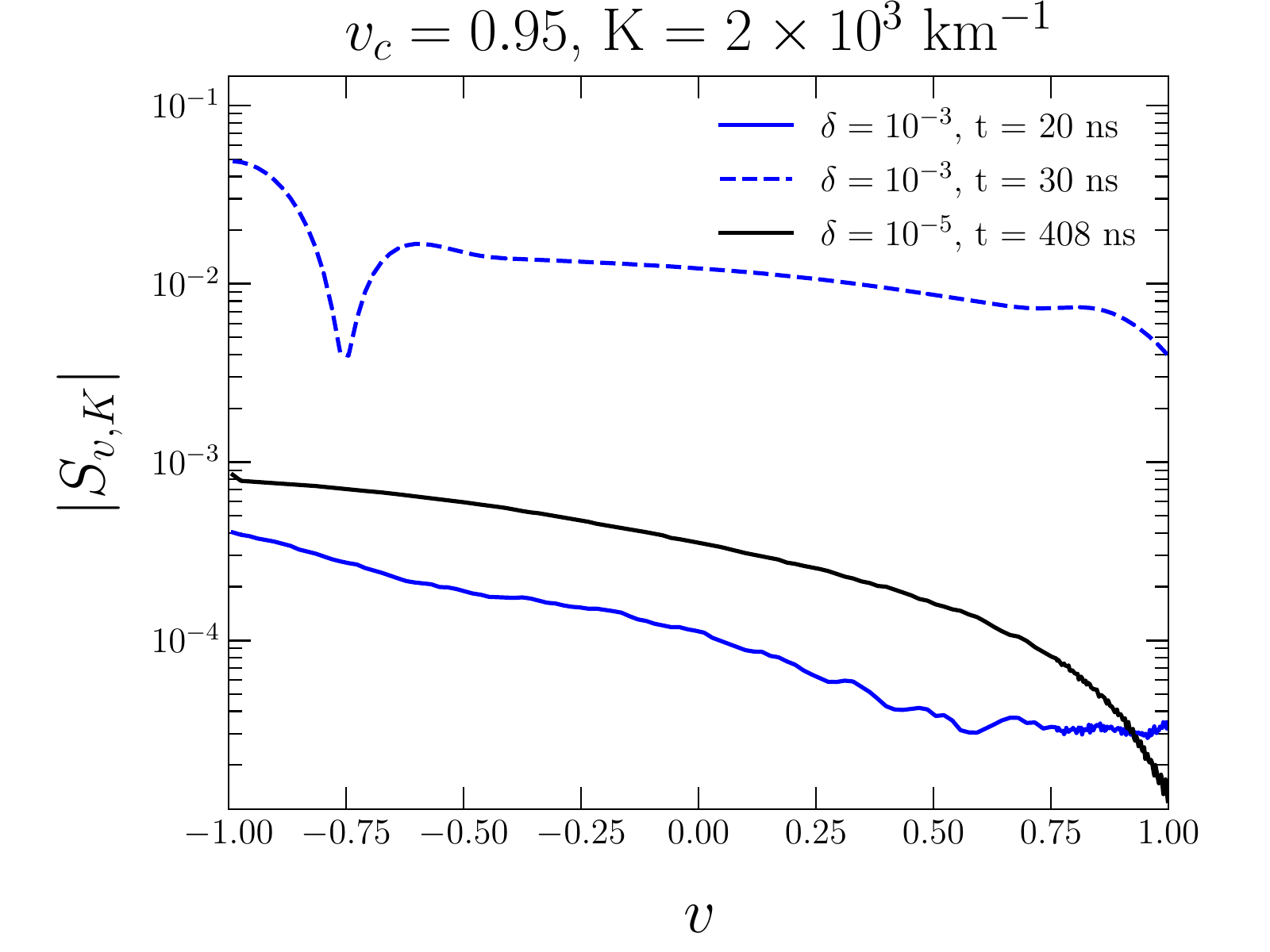}
\end{center}
\caption{
The angular distributions of the  eigenvectors 
of unstable modes, $S_{v,K}$ (Eq.~\eqref{eq:Sv_numerical}),   for $v_c=0.95$, and $\delta=10^{-3}$ and $10^{-5}$.
The initial perturbations are flat in $v$.}
\label{fig:non_linear_1}
\end{figure*}

\begin{figure*}[tb!] 
\centering
\begin{center}
\includegraphics[width=0.49\textwidth]{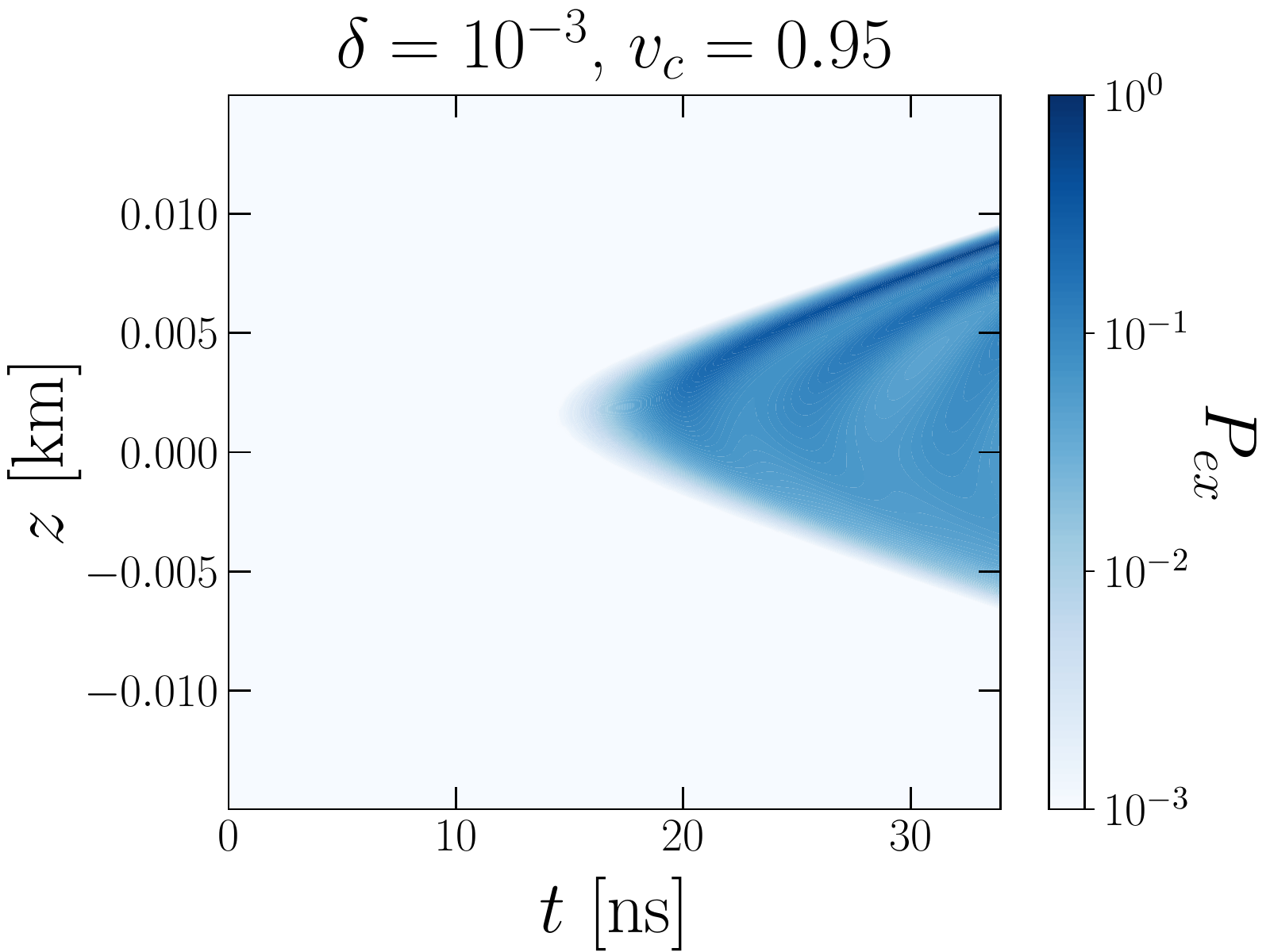}
\includegraphics[width=0.49\textwidth]{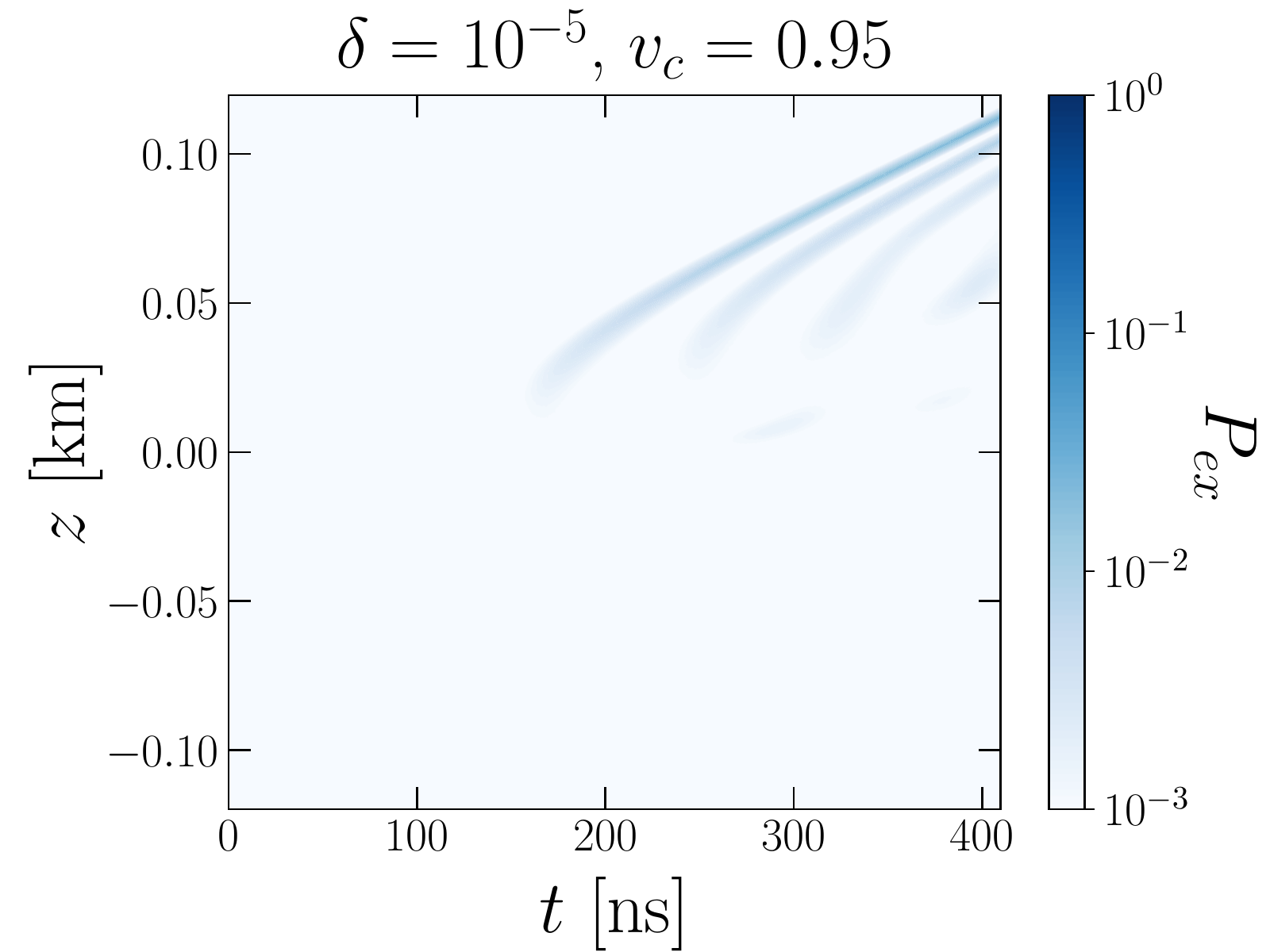}
\end{center}
\caption{
Time and space evolution of the conversion probability between the $e$ and $x$ flavors for $\delta=10^{-3}$ (left panel) and $10^{-5}$ (right panel).}
\label{fig:non_linear_2}
\end{figure*}

In order to extract the eigenvectors of unstable modes $Q_v$, we first calculate $S_v(t,z)$ as 
\begin{equation}
S_v(t,z)=\frac{\varrho_{ex}(t,z)}{\varrho_{ee}(t=0,z)-\varrho_{xx}(t=0,z)}
\label{eq:Sv_numerical}
\end{equation}
and then we perform its Fourier Transform 
\begin{equation}
S_{v,K}(t)=\frac{1}{L}\int_{0}^{L}\,dz\, S_v(t,z)\, e^{i K z},
\label{eq:Sv_numerical}
\end{equation}
$S_{v,K}(t)$ then represents an estimate of the eigenvectors $Q_v$ calculated in the linear regime. 

Here we focus on the cases of $\delta=10^{-3}$ and $10^{-5}$, and $v_c=0.95$. Similar results are obtained for $v_c=0.8$. 
Fig.~\ref{fig:non_linear_1} shows the angular distributions of $S_{v,K}(t)$ at different times for $K = 2\times 10^3$ km$^{-1}$. We note that the shape of $S_{v,K}$ is similar to that of $Q_v$ reported in Fig.~\ref{fig:linear}, as expected. 
Moreover,  we have checked that
we are already in the nonlinear regime and  thus $|S_{v,K}|$ is not growing exponentially anymore.

Fig.~\ref{fig:non_linear_2} displays the time and space evolution of the conversion probability between $e$ and $x$ flavor, defined as
\begin{equation}
P_{ex}(t,z)=1-\frac{\langle\varrho_{ee}(t,z)\rangle}{\langle\varrho_{ee}(0,z)\rangle},
\label{eq:Pex}
\end{equation}
where $\langle\varrho_{ee}(t,z)\rangle$ is the angle integrated equivalent of $\varrho_{ee}$,
\begin{equation}
\langle\varrho_{ee}(t,z)\rangle=\int_{-1}^1\,\mathrm{d}v\,\varrho_{v,ee}(t,z)\,.
\label{eq:rhoxx_numerical}
\end{equation}

While large flavor conversions can exist for the case of $\delta=10^{-3}$, they are significantly suppressed 
for $\delta=10^{-5}$ with the maximum of $P_{ex}$ being $\mathcal{O}(10^{-2})$,
despite the presence of fast flavor instabilities with large growth rates.
This can be understood by noting that for the $\delta=10^{-5}$ case, $|Q_{\rm{forward}}|/|Q_{\rm{backward}}| \lesssim 10^{-2}$.
Therefore, the amount of flavor conversion of the forward propagating neutrinos
is remarkably  small at the onset of the nonlinear regime. 
Then as the backward traveling neutrinos enters the nonlinear regime, the exponential growth ceases and 
the conversion of forward traveling neutrinos is suppressed. 
Nevertheless, 
the reason for observing  significant conversions
for the case of $\delta=10^{-3}$ is twofold. First of all, the population of the backward traveling neutrinos is relatively large
in this case, namely $\sim 5\%$ of the total neutrinos.   In addition and as a consequence of this,
$|Q_{\rm{forward}}|/|Q_{\rm{backward}}| \gtrsim 10\%$. Hence, 
 the forward traveling neutrinos can experience significant conversions even at the onset of the nonlinear regime.
 
 This result is indeed consistent
with the observation of Ref.~\cite{Zaizen:2021wwl}, where realistic ELN distributions of Eq.~\eqref{Eq:fnu} 
were employed. As can be seen in the lower right panel of Fig.~\ref{fig:linear}, $|Q_{\rm{forward}}|/|Q_{\rm{backward}}| \lesssim 10^{-3}$
for the realistic distributions.
Note that these results are obtained  in the absence of slow modes ($\omega=0$)
and one can still observe non-negligible flavor conversions in the presence of slow modes as in Ref.~\cite{Zaizen:2021wwl}, which has then
little to do with the backward ELN crossings and fast modes. 

It is also illuminating to note that while for $\delta=10^{-3}$ the flavor conversion wave travels both in the backward
and forward directions, it mostly affects the forward regions in the $\delta=10^{-5}$ case. This simply comes from the fact 
that for latter, there is a smaller number of neutrinos traveling in the backward direction. We have confirmed that there 
is absolutely no backward traveling conversion wave if one completely removes the backward traveling neutrinos, namely
 $f_\nu = 0$ for $v<0$.

Considering the linear stability analysis, our results suggest that the transition between significant and negligible flavor conversions occurs for
$\delta \lesssim 10^{-4}$, as a very rough estimate. Needless to say, the precise transition value should depend
on the exact angular distribution and the employed neutrino gas model.

\section{Conclusions}

Fast flavor conversions of SN neutrinos can happen on timescales as short as a few nanoseconds. 
A very generic class of  fast instabilities is the scattering-induced ones which are caused by the backward ELN crossings.
They actually occur
 due to the residual inelastic collisions of neutrinos on nucleons and/or nuclei  when they are nearly free-streaming.
 Such instabilities  can appear ubiquitously  both in the post-shock and pre-shock SN regions at large distances from the SN core .

Despite their ubiquity in the SN environment, we show that the scattering-induced fast instabilities
can not lead to significant flavor conversions if the population of the backward traveling neutrinos is not large enough. 
This arises from the fact that the unstable 
flavor states of  scattering-induced fast instabilities are extremely peaked at backward angles,
meaning that the instability should mainly impact the small fraction of neutrinos traveling in the backward direction.
This is quite understandable since  backward ELN crossings are themselves generated by that tiny 
fraction of neutrinos traveling in the backward direction.

Though our study is performed  in the two-flavor scenario, one should expect similar results in the three-flavor scenario. This can be simply understood by noting that the shape of the unstable eigenfunctions should be similar in the three-flavor scenario where analogous angular crossings can happen in all the three sectors ($e\mu$, $e\tau$ and $\mu\tau$)~\cite{Capozzi:2020syn}.

The results we obtained here suggest that one can safely ignore the scattering-induced fast instabilities in the SN environment
when the population of the backward traveling neutrinos is~$\ll 1\%$~\footnote{Unless  the tiny population of the backward traveling neutrinos  can lead to some unexpected non-negligible effects.}.
Two important questions still remain to be answered. First and foremost, it is not very clear to us if one can always assume that
 for all SN zones where scattering-induced crossings can occur, the depth of the crossing is such small. 
  If one can find scattering-induced ELN crossings
 for which the population of the backward traveling neutrinos is $\sim$ a few percents, then the next question would be
how the flavor content of the neutrino gas is affected by such instabilities. 
This is specially interesting given the fact that  the scattering-induced fast instabilities seem to be most generic/ubiquitous
class of fast instabilities in the SN environment.

\acknowledgments
We are  grateful to Georg Raffelt for helpful discussions and his comments on the manuscript. 
SA acknowledges support by the Deutsche Forschungsgemeinschaft through
Sonderforschungbereich SFB 1258 Neutrinos and Dark Matter in Astro- and
Particle Physics (NDM).
The work of FC at Virginia Tech is supported by the U.S. Department of Energy under the award number DE-SC0020250. The work of FC at IFIC is supported by GVA Grant No. CDEIGENT/2020/003.



\bibliographystyle{elsarticle-num}
\bibliography{Biblio}



\end{document}